\documentclass[11pt]{article}
\usepackage{amssymb}
\usepackage{graphicx}

\def\D{{\cal D}}

\def\H{{\cal H}}
\def\Hih{{\cal H}_{\rm HI}}

\def\S{{\Sigma}}

\def\SU{{\rm SU}}
\def\U{{\rm U}}

\def\lp{\ell_{\rm Pl}}

\def\be{\begin{equation}}
\def\ee{\end{equation}}
\def\ba{\begin{eqnarray}}
\def\ea{\end{eqnarray}}

\newcommand{\address}[1]{\vbox{\let\\=\cr \normalsize \vskip 1em
   \lineskip\normallineskip \halign{\hfil##\hfil\crcr#1\crcr}}}

\begin{document}

\title{Quantum Geometry In Action: Big Bang and Black Holes}
\author{Abhay Ashtekar 
 \\
\address{Center for Gravitational Physics and Geometry, \\
Physics Department, Penn State, University Park, PA 16802, USA}}

\maketitle
\begin{abstract}
Over the last three years, a number of fundamental issues in
quantum gravity were addressed in the framework of quantum
geometry, discussed extensively by John Baez in this conference.
In particular, these include: A statistical mechanical derivation
of the horizon entropy, encompassing astrophysically interesting
black holes cosmological horizons, and a natural resolution of the
big-bang singularity. The goal of this article is to communicate
these advances in general terms.

\end{abstract}

\section{Introduction}
\label{s1}

Major paradigm shifts in theoretical physics have required
mathematical arenas that were, at the time, new to physics.
Newton's mechanics and theory of gravitation could not have been
introduced without calculus; Maxwell's electrodynamics required
partial differential equations and analysis; Einstein had to learn
differential geometry to develop general relativity; and quantum
mechanics needed the theory of Hilbert spaces and operator
algebras.  It is widely believed that quantum gravity will lead to
the next profound paradigm shift in physics. What would be the
required mathematical arenas?  The answers to this question vary.
For example, Roger Penrose's twistor theory posits that space-time
would be a secondary, derived concept, arising from a
4-dimensional \textit{complex} space; the fundamental theory would
be based on complex manifolds, sheaf-cohomology and algebraic
geometry. Alain Conne's approach aims at describing fundamental
physics through non-commutative geometry.  In the loop quantum
gravity approach I will discuss here, the basic tool is Riemannian
quantum geometry. Just as differential geometry provides the
mathematical language to formulate classical gravitational
theories, such as general relativity, a specific quantum
Riemannian geometry provides the required setting for quantum
gravitational theories. Since this subject was covered in detail
by Baez, I will only provide a semi-qualitative introduction and
focus, rather, on applications of this quantum geometry. However,
let me make a general remark in this Introduction. As Dennis
Sullivan emphasized during discussions at the conference, from the
perspective of graph theory, freedom in the construction of a
(background independent) quantum theory of geometry is very
limited. Thus, the mathematical structures, definitions and
constructions we use are not only natural from this perspective,
but essentially unique.

Let us now turn to physics. What are some of the central
\textit{physical} and conceptual questions of quantum gravity? I
would like to outline a few of these to give a flavor of the
subject. Since this article is addressed to non-specialists, I
will select questions that arise from what we already know about
Nature, and what we expect based on physical theories which are
\textit{firmly} based on observations, avoiding issues
---such as higher dimensions and supersymmetry-- that are internal
to specific quantum gravity programs and which have yet to receive
observational support. Further discussion of the background
material can be found in Section \ref{s3}.

$\bullet$ \textit{Big-Bang and other singularities:} It is widely
believed that the prediction of a singularity, such as the
big-bang of classical general relativity, is primarily a signal
that the physical theory has been pushed beyond the domain of its
validity. A key question to any quantum gravity theory, then, is:
What replaces the big-bang? Is there a mathematically consistent
description of the evolution of the quantum state of the universe
which is singularity free?  General relativity predicts that the
space-time curvature must grow as we approach the big-bang but we
expect the quantum effects, ignored by general relativity, to
intervene, making quantum gravity indispensable before infinite
curvatures are reached. If so, what is the upper bound on
curvature? How close to the big-bang can we `trust'  classical
general relativity? What can we say about the `initial
conditions', i.e., the quantum state of geometry and matter that
correctly describes the big-bang? If they have to be imposed
externally, is there a \textit{physical} guiding principle?

$\bullet$ \textit{Black holes:} In the early seventies, using
imaginative thought experiments, Jacob Bekenstein argued that
black holes must carry entropy proportional to their area. About
the same time, Jim Bardeen, Brandon Carter and Stephen Hawking
(BCH) showed that black holes in equilibrium obey two basic laws,
which have the same form as the zeroth and the first laws of
ordinary thermodynamics, provided one replaces the area of the
black hole horizon $a_{\rm hor}$ by a multiple of the entropy $S$
in thermodynamics and black hole surface gravity $\kappa$ by a
corresponding multiple of the temperature $T$.\relax
\footnote{One can think of the horizon as the `surface' of the
black hole. In classical general relativity, one can not send
causal signals from the region within the horizon to the region
outside. Surface gravity $\kappa$ is, roughly, the black hole
analog of the acceleration $g$ due to gravity on the surface of
the earth.}
However, at first this similarity was thought to be only a formal
analogy because the BCH analysis was based on \textit{classical}
general relativity and simple dimensional considerations show that
the proportionality factors must involve Planck's constant
$\hbar$. Two years later, using quantum field theory on a black
hole background space-time, Hawking showed that black holes in
fact radiate quantum mechanically as though they are black bodies
at temperature $T = \hbar\kappa/2\pi$. Using the analogy with the
first law, one can then conclude that the black hole entropy
should be given by $S_{\rm BH} = a_{\rm hor}/4G\hbar$, where $G$
is Newton's gravitational constant. This conclusion is striking
and deep because it brings together the three pillars of
fundamental physics
---general relativity, quantum theory and statistical mechanics.
However, the argument itself is a rather hodge-podge mixture of
classical and semi-classical ideas, reminiscent of the Bohr theory
of atom. A natural question then is: what is the analog of the
more fundamental, Pauli-Schr\"odinger theory of the Hydrogen atom?
More precisely, what is the statistical mechanical origin of black
hole entropy? What is the nature of a quantum black hole and what
is the interplay between the quantum degrees of freedom
responsible for entropy and the exterior curved geometry? Can one
derive the Hawking effect from first principles of quantum
gravity?

$\bullet$ \textit{Planck scale physics and the low energy world:}
Perhaps the central lesson of general relativity is that
\textit{gravity is geometry}. There is no longer a background
metric, no inert stage on which dynamics unfolds. Geometry itself
is dynamical. Therefore, one expects that a fully satisfactory
quantum gravity theory would also be free of a background
space-time geometry. However, of necessity, a background
independent description must use physical concepts and
mathematical tools that are quite different from those of the
familiar, low energy physics formulated in flat space-time. A
major challenge then is to show that this low energy description
does arise from the pristine, Planckian world in an appropriate
sense.\relax
\footnote{The characteristic length scale in the Planck regime is
$\sim \, 10^{-33}{\rm cm}$ while the smallest distance we can
probe with our highest energy accelerators today is $\sim \,
10^{-17} {\rm cm}$.}
In this `top-down' approach, does the fundamental theory admit a
`sufficient number' of semi-classical states? Do these
semi-classical sectors provide enough of a background geometry to
anchor low energy physics? can one recover the familiar
description? Furthermore, can one pin point why the standard
`bottom-up' perturbative approach fails? That is, what is the
essential feature which makes the fundamental description
mathematically coherent, but is absent in the standard
perturbative quantum gravity?

Of course, this is by no means a complete list of challenges.
There are many others. Since there is no background space-time
metric, what does `time evolution' mean?  Without a fixed
space-time at one's disposal, what is one to make of quantum
measurement theory and the associated questions of interpretation
of quantum mechanics? What role does space-time topology play and
can it change? \ldots. Recent advances within loop quantum gravity
have led to illuminating answers to many of these questions and
opened-up avenues to address others. As in my talk, in this
report, I will focus on issues related to black holes and
big-bang.

\section{A bird's eye view of loop quantum gravity}
\label{s2}

In this section, I will briefly summarize the salient features and
current status of quantum geometry. The emphasis is on structural
and conceptual issues; details can be found in references [1-9].

\subsection{Viewpoint}
\label{s2.1}

In this approach, one takes the central lesson of general
relativity seriously: gravity \textit{is} geometry whence, in a
fundamental theory, there should be no background metric. In
quantum gravity, geometry and matter should \textit{both} be `born
quantum mechanically'. Thus, in contrast to approaches developed
by particle physicists, one does not begin with quantum matter on
a background geometry and use perturbation theory to incorporate
quantum effects of gravity. There \textit{is} a manifold but no
metric, or indeed any other physical fields, in the background.%
\footnote{In 2+1 dimensions, although one begins in a completely
analogous fashion, in the final picture one can get rid of the
background manifold as well. Thus, the fundamental theory can be
formulated \textit{combinatorially} \cite{loop,books}. To achieve
this goal in 3+1 dimensions, one needs a much better understanding
of the theory of (intersecting) knots in 3 dimensions.}
In classical gravity, Riemannian geometry provides the appropriate
mathematical language to formulate the physical, kinematical
notions as well as the final dynamical equations. This role is now
taken by \textit{quantum} Riemannian geometry, discussed below. In
the classical domain, general relativity stands out as the best
available theory of gravity, some of whose predictions have been
tested to an amazing accuracy, surpassing even the legendary tests
of quantum electrodynamics. However, if one applies to general
relativity the standard perturbative techniques of quantum field
theory, one obtains a `non-renormalizable' theory , i.e., a theory
with uncontrollable infinities. Therefore, it is natural to ask:
\textit{Does quantum general relativity, coupled to suitable
matter} (or supergravity, its supersymmetric generalizations)
\textit{exist as a consistent theory non-perturbatively ?} There
is no a priori implication that such a theory would be the final,
complete description of Nature. Nonetheless, this is a fascinating
open question at the level of mathematical physics.

In the particle physics circles, the answer is often assumed to be
in the negative, not because there is concrete evidence against
non-perturbative quantum gravity, but because of an analogy to the
theory of weak interactions, where non-renomalizability of the
initial `Fermi theory' forced one to replace it by the
renormalizable Glashow-Weinberg-Salam theory. However this analogy
overlooks the crucial fact that, in the case of general
relativity, there is a qualitatively new element. Perturbative
treatments pre-suppose that the space-time can be assumed to be a
continuum \textit{at all scales} of interest to physics under
consideration. Since this is a safe assumption for weak
interactions, non-renormalizability was a genuine problem.
However, in the gravitational case, the scale of interest is given
by the Planck length $\lp$ and there is no physical basis to
pre-suppose that the continuum picture should be valid down to
that scale. The failure of the standard perturbative treatments
may simply be due to this grossly incorrect assumption and a
non-perturbative treatment which correctly incorporates the
physical micro-structure of geometry may well be free of these
inconsistencies.

As indicated above, even if quantum general relativity did exist
as a mathematically consistent theory, there is no a priori reason
to assume that it would be the `final' theory of all known
physics. In  classical general relativity, while requirements of
background independence and general covariance do restrict the
form of interactions between gravity and matter fields and among
matter fields themselves, they do not \textit{determine} these
interactions. Quantum general relativity would have the same
limitation. Put differently, such a theory would not be a
satisfactory candidate for unification of all known forces.
However, just as general relativity has had powerful implications
in spite of this limitation in the classical domain, quantum
general relativity should have qualitatively new predictions,
pushing further the existing frontiers of physics. Indeed,
unification does not appear to be an essential criterion for
usefulness of a theory even in other interactions. Quantum
chrmodynamics (QCD) for example, is a powerful theory of strong
interactions  even though it does not unify them with electro-weak
ones. Furthermore, the fact that we do not yet have a viable
candidate for the grand unified theory does not make QCD any less
useful.

\subsection{Quantum Geometry}
\label{s2.2}

Although there is no natural unification of dynamics of all
interactions in loop quantum gravity, it does provide a
kinematical unification. More precisely, in this approach one
begins by formulating general relativity in the mathematical
language of {connections}, the basic variables of gauge theories
of electro-weak and strong interactions. Thus, now the
configuration variables are not metrics (as in Wheeler's
geometrodynamics program), but certain \textit{spin connections};
the emphasis is shifted from distances to holonomies
\cite{class,books}. Consequently, the basic kinematical structures
are the same as those used in gauge theories. A key difference,
however, is that while a background space-time metric is available
and crucially used in gauge theories, now there are no background
fields whatsoever. This absence is forced on us by the requirement
of diffeomorphism invariance.

This is a key difference and it causes a host of conceptual as
well as technical difficulties in the passage to quantum theory.
For, most of the techniques used in the familiar, Minkowskian
quantum theories are deeply rooted in the availability of a flat
back-ground metric. It is this structure that enables one to
single out the vacuum state, perform Fourier transforms to
decompose fields canonically in to creation and annihilation
operators, define masses and spins of particles and carry out
regularizations of products of operators. Already when one passes
to quantum field theory in curved space-times, extra work is
needed to construct mathematical structures that are adequate for
physics. In our case, the situation is much more drastic: there is
no background metric what so ever. Therefore new physical ideas
and mathematical tools are now necessary. Fortunately, they were
constructed by a number of researchers in the mid-nineties and
have given rise to a detailed quantum theory of geometry
\cite{qg1,qg2,qg3,bi,qg4}.

Because the situation is conceptually so novel and because there
are no direct experiments to guide us, reliable results require
mathematical precision to ensure that there are no hidden
infinities.  Achieving this precision has been a high priority in
the program. Thus, while one is inevitably motivated by heuristic,
physical ideas and formal manipulations, the final results are
mathematically rigorous. In particular, due care is taken in
constructing function spaces, defining measures and functional
integrals, regularizing products of field operator, and
calculating eigenvectors and eigenvalues of geometric operators.
The final results are all free of divergences, well-defined, and
respect the background independence and diffeomorphism invariance.

Let me now turn to specifics. It is perhaps simplest to begin with
a Hamiltonian or symplectic description of general relativity. The
phase space is the cotangent bundle. The configuration variable is
a connection, $A$ on a fixed 3-manifold $\S$ representing `space'
and (as in gauge theories) the momenta are the `electric field'
2-forms $E$, both of which take values in the Lie-algebra of
$\SU(2)$. In the present gravitational context, the momenta
acquire a geometrical significance: their Hodge-duals ${}^\star\!
E$ can be naturally interpreted as orthonormal triads (with
density weight $1$) and determine the dynamical, Riemannian
geometry of $\S$. Thus, (in contrast to Wheeler's
geometrodynamics) the Riemannian structures on $\S$ are now built
from \textit{momentum} variables. The basic kinematic objects are
holonomies of $A$, which dictate how spinors are parallel
transported along curves, and the 2-forms $E$, which determine the
Riemannian metric of $\S$. (Matter couplings to gravity have also
been studied extensively \cite{class,books}.)

In the quantum theory, the fundamental excitations of geometry are
most conveniently expressed in terms of holonomies
\cite{loop,qg1}. They are thus one-dimensional, polymer-like and,
in analogy with gauge theories, can be thought of as `flux lines
of the electric field'. More precisely, they turn out to be flux
lines of areas: an elementary flux line deposits a quantum of area
on any 2-surface $S$ it intersects. Thus, if quantum geometry were
to be excited along just a few flux lines, most surfaces would
have zero area and the quantum state would not at all resemble a
classical geometry. Semi-classical geometries can result only if a
huge number of these elementary excitations are superposed in
suitably dense configurations \cite{sc,fock}. The state of quantum
geometry around you, for example, must have so many elementary
excitations that $\sim 10^{68}$ of them intersect the sheet of
paper you are reading, to endow it an area of $\sim 100 {\rm
cm}^2$. Even in such states, the geometry is still distributional,
concentrated on the underlying elementary flux lines; but if
suitably coarse-grained, it can be approximated by a smooth
metric. Thus, the continuum picture is only an approximation that
arises from coarse graining of semi-classical states.

These quantum states span a specific Hilbert space $\H = L^2
(\bar{\cal A}, d\mu_o)$, consisting of functions on the space of
(suitably generalized) connections which are square integrable
with respect to a natural, diffeomorphism invariant (regular,
Borel) measure $\mu_o$ \cite{qg1}. This space is very large.
However, it can be conveniently decomposed in to a family of
orthonormal, \textit{finite} dimensional sub-spaces $\H =
\oplus_{\gamma, \vec{j}} \H_{\gamma, \vec{j}}$, labelled by finite
graphs $\gamma$ each edge of which itself is labelled by a
non-trivial irreducible representation of $\SU(2)$ (or, a
half-integer,  or a spin ${j}$) \cite{qg2}. $\H_{\gamma,\vec{j}}$
can be regarded as the Hilbert space of a `spin-system'. These
spaces are extremely simple to work with; this is why very
explicit calculations are feasible. Elements of
$\H_{\gamma,\vec{j}}$ are referred to as \textit{spin-network
states} \cite{qg2}.

As one would expect from the structure of the classical theory,
the basic quantum operators are the holonomies $\hat{h}_p$ along
paths $p$ in $\S$ and the triads $\widehat{{}^\star\! E}$
\cite{qg3}. Both sets of operators are densely defined and
self-adjoint on $\H$. Furthermore, a striking result is that
\textit{all eigenvalues of the triad operators are discrete.} This
key property is, in essence, the origin of the fundamental
discreteness of quantum geometry. For, just as the classical
Riemannian geometry of $\S$ is determined by the triads ${}^\star
\! E$, all Riemannian geometry operators ---such as the area
operator $\hat{A}_S$ associated with a 2-surface $S$ or the volume
operator $\hat{V}_R$ associated with a region $R$--- are
constructed from $\widehat{{}^\star\! E}$. However, since even the
classical quantities $A_S$ and $V_R$ are non-polynomial
functionals of the triads, the construction of the corresponding
$\hat{A}_S$ and $\hat{V}_R$ is quite subtle and requires a great
deal of care. But their final expressions are rather simple
\cite{qg3}.

In this regularization, the underlying background independence
turns out to be a blessing. For, diffeomorphism invariance
constrains the possible forms of the final expressions
\textit{severely} and the detailed calculations then serve
essentially to fix numerical coefficients and other details. Let
us illustrate this point with the example of the area operators
$\hat{A}_S$. Since they are associated with 2-surfaces $S$ while
the states have 1-dimensional support, the diffeomorphism
covariance requires that the action of $\hat{A}_S$ on a state
$\Psi_{\gamma, \vec{j}}$ must be concentrated at the intersections
of $S$ with $\gamma$. The detailed expression bears out this fact:
the action of $\hat{A}_S$ on $\Psi_{\gamma, \vec{j}}$ is dictated
simply by the spin labels $j_I$ attached to those edges of
$\gamma$ which intersect $S$. For all surfaces $S$ and
3-dimensional regions $R$ in $\S$,  $\hat{A}_S$ and $\hat{V}_R$
are densely defined, self-adjoint operators. \textit{All their
eigenvalues are discrete} \cite{qg3}. Naively, one might expect
that the eigenvalues would be uniformly spaced, given by, e.g.,
integral multiples of the Planck area or volume.  This turns out
\textit{not} to be the case; the distribution of eigenvalues is
quite subtle. In particular, the eigenvalues crowd rapidly as
areas and volumes increase. In the case of area operators, the
complete spectrum is known in a closed form, and the first several
hundred eigenvalues have been explicitly computed numerically. For
a large eigenvalue $a_n$, the separation $\Delta a_n = a_{n+1}
-a_n$ between consecutive eigenvalues decreases exponentially:
$\Delta a_n \, \le\, \lp^2 \, \exp -(\sqrt{a_n}/\lp )$! Because of
such strong crowding, the continuum approximation becomes
excellent quite rapidly just a few orders of magnitude above the
Planck scale. At the Planck scale, however, there is a precise and
very specific replacement. This is the arena of quantum geometry.
The premise is that the standard perturbation theory fails because
it ignores this fundamental discreteness (see Section \ref{s2.1}).

There is however a key mathematical subtlety \cite{class,bi}. This
non-perturbative quantization has a one parameter family of
ambiguities labelled by $\gamma > 0$. This $\gamma$ is called the
Barbero-Immirzi parameter (and is rather similar to the well-known
$\theta$-parameter of QCD). In the classical theory, $\gamma$ is
irrelevant but in quantum theory different values of $\gamma$
correspond to unitarily inequivalent representations of the
algebra of geometric operators. The overall mathematical structure
of all these sectors is very similar; the only difference is that
the eigenvalues of all geometric operators scale with $\gamma$.
For example, the simplest eigenvalues of the area operator
$\hat{A}_S$ in the $\gamma$ quantum sector is given by
\be \label{2.1} a_{\{j\}} = 8\pi\gamma \lp^2 \, \sum_I \sqrt{j_I
(j_I +1)} \ee
where $I = 1,\ldots N$ for some integer $N$ and each $j_I$ is a
half-integer. Since the representations are unitarily
inequivalent, as usual, one must rely on Nature to resolve this
ambiguity: Just as Nature must select a specific value of $\theta$
on QCD, it must select a specific value of $\gamma$ in loop
quantum gravity. With one judicious experiment ---e.g.,
measurement of the lowest eigenvalue of the area operator
$\hat{A}_S$ for a 2-surface $S$ of any given topology--- we could
determine the value of $\gamma$ and fix the theory. Unfortunately,
such experiments are hard to perform! However, we will see in
Section \ref{s3.2} that the Bekenstein-Hawking formula of black
hole entropy provides an indirect measurement of this lowest
eigenvalue of area for the 2-sphere topology and can therefore be
used to fix the value of $\gamma$.

\section{Applications of quantum geometry}
\label{s3}

In this section, I will summarize two recent developments that
answer several of the questions raised under first two bullets in
the Introduction.

\subsection{Big bang}
\label{s3.1}

Let us first recall how the big-bang singularity arises in
classical general relativity. Observations have shown that the
universe is spatially homogeneous and isotropic on a cosmological
scale. Therefore, to model the large scale behavior of the
universe, one begins by assuming that the 4-manifold representing
space-time is foliated by 3-dimensional spatial manifolds, each
equipped with a metric of constant curvature. Thus there are three
possibilities: the leaves of the preferred foliation are either
metric 3-spheres, or flat, or metric 3-hyperboloids, depending on
whether the scalar curvature (which is constant on each leaf) is
positive, zero or negative. To be specific, let me assume the
first case. Then, although the scalar curvature is constant on any
one spatial slice, it changes in time, giving rise to an overall
expansion or contraction. The radius $a$ of the 3-sphere encodes
the full information of the 3-metric at that instant of time and
is called the \textit{scale factor}. One then suitably models
matter-sources ---galaxies and the observed radiation fields---
and seeks solutions of Einstein's equation with these symmetries.
The equation implies that the universe must have `originated from
a big-bang': if we evolve the solution backwards in time, the
scale factor $a$ must eventually go to zero and the curvature must
diverge as $1/a^2$. At this `initial instant', Einstein's equation
breaks down; classical physics stops. As discussed in the
Introduction, the general belief is that this singular behavior is
an artifact of our insistence of applying general relativity
beyond the domain of its validity. Quantum effects are thought to
intervene and dominate the `real physics' in the high curvature
regions. The question then is: what replaces the big-bang in this
new, more accurate theory?

This question has been discussed for over thirty years in a
framework called `quantum cosmology'. Traditionally, one has
proceeded by first imposing spatial symmetries
---such as homogeneity and isotropy--- to freeze out all but a
finite number of degrees of freedom \textit{already at the
classical level} and then quantizing the reduced system. In the
simplest case, the basic variables of the reduced classical system
are the scale factor $a$ and matter fields $\phi$. One then asks:
in the theory so quantized, do the singularities of the classical
general relativity disappear? Unfortunately, without an additional
input, they do not: typically, to resolve the singularity one
either had to introduce matter with unphysical properties or
introduce boundary conditions by invoking new principles.

In a series of seminal papers \cite{bb}, Martin Bojowald has shown
that the situation in loop quantum cosmology is quite different:
the underlying quantum geometry makes a \textit{qualitative}
difference very near the big-bang and naturally resolves the
singularity. In the standard procedure summarized above, the
reduction is carried out at the classical level and this removes
all traces of the fundamental discreteness. Therefore, the key
idea in Bojowald's analysis is to retain the essential features of
quantum geometry by first quantizing the kinematics of the
\textit{full theory} as in Section \ref{s2.2} and then restricting
oneself to \textit{quantum} states which are spatially homogeneous
and isotropic. As a result, the scale factor operator $\hat{a}$
has \textit{discrete eigenvalues}. The continuum limit is reached
rapidly. For example, the gap between an eigenvalue of $\hat{a}$
of $\sim 1 {\rm cm}$ and the next one is less than $\sim
10^{-30}\, \lp$! Nonetheless, near $a \sim \lp$ there are
surprises. Predictions of loop quantum cosmology are very
different from those of traditional quantum cosmology.

The first surprise occurs already at the kinematical level. Recall
that, in the classical theory curvature is essentially given by
$1/a^2$, and blows up at the big-bang. What is the situation in
quantum theory? Denote the Hilbert space of spatially homogeneous,
isotropic kinematical quantum states by $\Hih$. A self-adjoint
operator $\widehat{\rm curv}$ corresponding to curvature can be
constructed on $\Hih$ and turns out to be \textit{bounded from
above}. This is very surprising because $\Hih$ admits an
eigenstate of the scale factor operator $\hat{a}$ with a discrete,
zero eigenvalue! At first, it may appear that this could happen
only by an artificial trick in the construction of $\widehat{\rm
curv}$ and that this quantization can not possibly be right
because it seems to represent a huge departure from the classical
relation $({\rm curv})\, a^2 =1$. However, these concerns turn out
to be misplaced. The procedure for constructing $\widehat{\rm
curv}$ is natural and, furthermore, descends from full quantum
theory.

Let us examine the properties of $\widehat{\rm curv}$. Its upper
bound $u_{\rm curv}$ is finite but absolutely huge:
\be \label{3.1} u_{\rm curv} \, \sim \,\frac{256}{81}\,
\frac{1}{\lp^2} \equiv \frac{256}{81}\, \frac{1}{G\hbar} \ee
or, about $10^{77}$ times the curvature at the horizon of a solar
mass black hole. The functional form of the upper bound is also
illuminating. Recall first the Pauli-Schr\"odinger treatment of
the hydrogen atom in non-relativistic quantum mechanics. Because
the Coulomb potential between the proton (nucleus of the atom) and
the electron diverges as $- 1/r$, in the classical theory the
energy is unbounded from below. However, thank to the Planck's
constant $\hbar$, in the quantum theory, we obtain a finite value,
$E_0 = -\, (me^4 /\hbar^2)$. Similarly, $u_{\rm curv}$ is finite
because $\hbar$ is non-zero and tends to the classical answer as
$\hbar$ tends to zero.

\begin{figure}
\begin{center}
\includegraphics[height=3in]{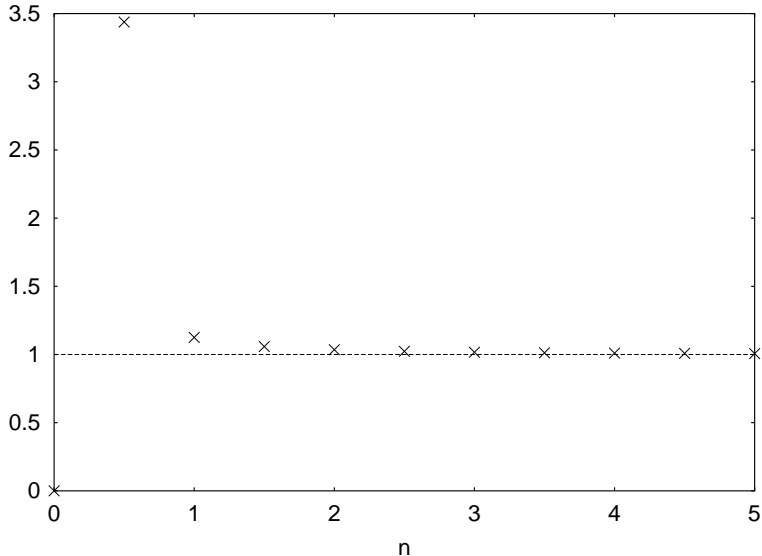}
\caption{\small{The product $a_n \cdot b_n $ as a function of $n$.
The corresponding classical product $a \cdot \sqrt{\rm curv}$
equals $1$. }}\label{prod}
\end{center}
\end{figure}

At curvatures as large as $u_{\rm curv}$, it is natural to expect
large departures from classical relations such as $({\rm curv})\,
a^2 =1$. But is this relation recovered in the semi-classical
regime? The answer is in the affirmative. In fact it is somewhat
surprising how quickly this happens. As one would expect, one can
simultaneously diagonalize $\hat{a}$ and $\widehat{\rm curv}$. If
we denote their eigenvalues by $a_n$ and $b_n$ respectively, then
$a_n\cdot b_n -1$ is of the order $10^{-4}$ at $n =100$ and
decreases rapidly as $n$ increases. These properties show that, in
spite of the initial surprise, the quantization procedure is
viable. Furthermore, one can apply it also to more familiar
systems such as a particle moving on a circle and obtain results
which at first seem surprising but are in complete agreement with
the standard quantum theory of these systems.

Since the curvature is bounded above in the entire Hilbert space,
one might hope that the quantum evolution may be well-defined
right through the big-bang singularity. Is this in fact the case?
The second surprise is that although the quantum evolution is
close to that of the so-called Wheeler-DeWitt equation of standard
quantum cosmology for large $a$, there are dramatic differences
near the big-bang which makes it well defined even \textit{at} the
big-bang, \textit{without any additional input}. To solve the
quantum Einstein equation, Bojowald again follows, step by step,
the procedure introduced (by Thomas Thiemann) in the full theory.
Let us expand the full quantum state as $\mid\Psi\!> = \sum_n
\psi_n (\phi) \mid n\!>$ where $\mid n\!>$ are the eigenstates of
the scale factor operator and $\phi$ denotes matter fields. Then,
the quantum Einstein equation takes the form:
\be \label{3.2} c_n \psi_{n+8}(\phi) + d_{n} \psi_{n+4} (\phi) +
e_n \psi_{n} (\phi)+ f_n \psi_{n-4} (\phi) + g_n \psi_{n-8}(\phi)
\, = \, \gamma \lp^2 \,\,\hat{H}_\phi \psi_n(\phi) \ee
where $c_n,\ldots g_n$ are fixed numerical coefficients, $\gamma$
the Barbero-Immirzi parameter and $\hat{H}_\phi$ is the matter
Hamiltonian. (Again, using the Thiemann regularization, one can
show that the matter Hamiltonian is a well-defined operator.)

As one would expect from the phase space-formulation of classical
general relativity, primarily, Eq(\ref{3.2}) serves to constrain
the coefficients $\psi_n(\phi)$ of physically permissible quantum
states. However, \textit{if} we choose to interpret the scale
factor (more precisely, the square of the scale factor times the
determinant of the triad) as a time variable, Eq(\ref{3.2}) can be
interpreted as an `evolution equation' which evolves the state
through discrete time steps. In a (large) neighborhood of the
big-bang singularity, this notion of time is viable. For the
choice of factor ordering used in the Thiemann regularization, one
can evolve in the past through $n=0$, i.e. right through the
classical singularity. Thus, the infinities predicted by the
classical theory at the big-bang are indeed artifacts of assuming
that the classical, continuum space-time approximation is valid
right up to the big-bang. In the quantum theory, the state can be
evolved through the big-bang without any difficulty. However, the
classical space-time description fails near the big-bang; quantum
evolution is well-defined but the classical space-time
`dissolves'.

The `evolution' equation (\ref{3.2}) has other interesting
features. To begin with, the space of solutions is 16 dimensional.
Can we single out a preferred solution by imposing a
\textit{physical} condition? One possibility is to impose a
\textit{pre-classicality} condition, i.e., to require that the
quantum state not oscillate rapidly from one step to the next at
\textit{late} times when we know our universe behaves classically.
Although this is an extra input, it is not a theoretical prejudice
about what should happen at (or near) the big-bang but an
observationally motivated condition that is clearly satisfied by
our universe. The coefficients $c_n,\ldots g_n$ of (\ref{3.2}) are
such that this condition singles out a solution uniquely. One can
ask what this state does at negative times, i.e., before the
big-bang. (Time becomes negative because triads flip orientation
on the `other side'.) Preliminary indications are that the state
does not become pre-classical there. If this is borne out by
detailed calculations, then the `big-bang' separates two regimes;
on `our' side, classical geometry is both meaningful and useful at
late times  while on the `other' side, it is not.\relax
\footnote{There is thus a qualitative similarity to the phenomenon
of phase transitions in magnets. `Our side' of the big-bang is
analogous to the ferro-magnetic phase (the role of the
`magnetization' mean field ---the vector pointing from the south
to the north pole of a ferro-magnet---  being played by the
classical geometry) and the `other side' is analogous to the
para-magnetic phase (where `magnetization' is no longer a useful
concept.}
Another interesting feature is that the standard Wheeler-DeWitt
equation is recovered if we take the limit $\gamma \to 0$ and $n
\to \infty$ such that the eigenvalues of $\hat{a}$ take on
continuous values. This is completely parallel to the limit we
often take to coarse grain the quantum description of a rigidly
spinning rotor to `wash out' discreteness in angular momentum
eigenvalues and arrive at the classically allowed continuous
angular momenta. From this perspective, then, one is led to say
that the most striking of the consequences of loop quantum gravity
are not seen in standard quantum cosmology because it `washes out'
the fundamental discreteness of quantum geometry.

Finally, the detailed calculations have revealed another
surprising feature. The fact that the quantum effects become
prominent near the big bang, completely invalidating the classical
predictions, is pleasing but not unexpected. However, prior to
these calculations, it was not clear how soon after the big-bang
one can start trusting semi-classical notions and calculations. It
would not have been surprising if we had to wait till the radius
of the universe became, say, a few million times the Planck
length. These calculations strongly suggest that few hundred
Planck lengths should suffice. This is fortunate because it is now
feasible to develop quantum numerical relativity; with
computational resources commonly available, grids with $(10^6)^3$
points are hopelessly large but one with $(100)^3$ points are
readily available.

\subsection{Black-holes}
\label{s3.2}

Loop quantum cosmology illuminates dynamical ramifications of
quantum geometry but within the context of mini-superspaces where
all but a finite number of degrees of freedom are frozen. In this
sub-section, I will discuss a complementary application where one
considers the full theory but probes consequences of quantum
geometry which are not sensitive to full quantum dynamics ---the
application of the framework to the problem of black hole entropy.
This discussion is based on joint work with Baez, Corichi and
Krasnov \cite{bh} which itself was motivated by earlier work of
Krasnov, Rovelli and others.

Let us begin with classical general relativity. Consider first the
simplest solution of Einstein's equation: a manifold which is
topologically $R^4$, equipped with a flat metric (of signature
-+++). It has the property that an observer near infinity can
receive a causal signal from \textit{any} point in the interior,
sometime along its infinite world-line. Thus, no part of
space-time is permanently hidden from infinity. However,
Einstein's equation admits solutions which do not share this
property. In such solutions, the portion of space-time which is
hidden from all asymptotic observers is called a \textit{black
hole region}. Physically, since gravity is attractive, the
space-time metric around dense, compact astrophysical objects is
such that the light cones in their vicinity are `bent towards the
object'. Since causal signals propagate with a speed less than or
equal to that of light, it is `harder for the information to leak
out' from their vicinity. A black hole region results if light
cones are bent so much that they are `tilted completely inwards';
i.e., no causal signal can leave the region of space-time in
question. A space-like surface representing an `instant of time'
intersects the boundary of the black-hole region in a 2-sphere. We
will refer to this 2-sphere boundary as the \textit{black hole
horizon}. (This terminology is not the standard one but is more
convenient for our purposes here.)

As explained in the Introduction, since mid-seventies, a key
question in the subject has been: What is the statistical
mechanical origin of the black hole entropy $S_{\rm BH} =
\textstyle{a_{\rm hor}/ 4\lp^2}$? What are the microscopic degrees
of freedom that account for this entropy? This relation implies
that a solar mass black hole must have $\sim (\exp 10^{77})$
quantum states, a number that is \textit{huge} even by the
standards of statistical mechanics. Where do all these states
reside? To answer these questions, in the early nineties John
Wheeler suggested the following heuristic picture, which he
christened `It from Bit'. Divide the black hole horizon in to
elementary cells, each with one Planck unit, $\lp^2$, of area  and
assign to each cell two microstates, or one `bit'. Then the total
number of states ${\cal N}$ is given by ${\cal N} = 2^n$ where $n
= ({a_{\rm hor}/ \lp^2})$ is the number of elementary cells,
whence entropy is given by $S = \ln {\cal N} \sim a_{\rm hor}$.
Thus, apart from a numerical coefficient, the entropy (`It') is
accounted for by assigning two states (`Bit') to each elementary
cell. This qualitative picture is simple and attractive. Therefore
it is natural to ask if it can be made precise. Can these
heuristic ideas be supported by a systematic analysis from first
principles? What is the rationale behind dividing the black hole
horizon in to elementary cells of unit Planck area? Why are there
exactly two quantum states associated with each cell? It turned
out that quantum geometry could supply answers to these questions
through a detailed analysis.\relax
\footnote{I should add, however, that this account does not follow
chronology. Black hole entropy was computed in quantum geometry
quite independently and the realization that the `It from Bit'
picture works so well was somewhat of a surprise.}

A systematic approach requires that we first specify the class of
black holes of interest. Since the entropy formula is expected to
hold unambiguously for black holes in equilibrium, most analyses
were confined to space-times with `eternal' black holes admitting
a global time-translation isometry, rather than the astrophysical
ones which result from a gravitational collapse. From a physical
viewpoint however, this assumption seems overly restrictive. After
all, in statistical mechanical calculations of entropy of ordinary
systems, one only has to assume that the given system is in
equilibrium, not the whole world. Therefore, it should suffice to
assume that the black hole itself is in equilibrium; the exterior
geometry should not be forced to be time-independent. Finally, it
has been known since the mid-seventies that the thermodynamical
considerations apply not only to black holes but also to so-called
`cosmological horizons'. A natural question is: Can these diverse
situations be treated in a single stroke? Within the quantum
geometry approach, the answer is in the affirmative. The idea that
the black hole (or the cosmological horizon) is itself in
equilibrium is captured by certain boundary conditions which
ensure that the horizon itself is \textit{isolated}, allowing
time-dependent space-time geometry and matter fields in the
exterior region. Entropy associated with an isolated horizon
refers to the family of observers in the exterior region for whom
the isolated horizon is a physical boundary that separates the
region which is accessible to them from the one which is not.
(This point is especially important for cosmological horizons
where, without reference to observers, one can not even define
these horizons.) States which contribute to this entropy are the
ones which can interact with the states in the exterior; in this
sense, they `reside' on the horizon.

\begin{figure}
\begin{center}
\includegraphics[height=2in]{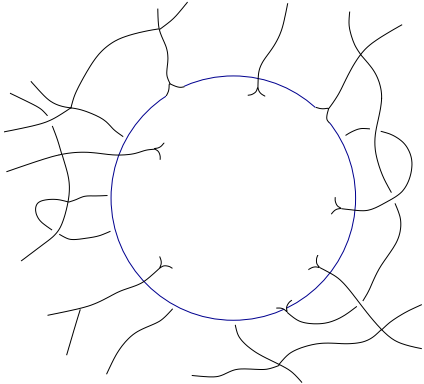}
\caption{\small{Quantum Horizon. Polymer excitations in the bulk
puncture the horizon, endowing it with quantized area.
Intrinsically, the horizon is flat except at punctures where it
acquires a quantized deficit angle. These angles add up to endow
the horizon with a 2-sphere topology.}} \label{fig2}
\end{center}
\end{figure}

In the detailed analysis, one considers space-times admitting an
isolated horizon as inner boundary and carries out a systematic
quantization. The quantum geometry framework can be naturally
extended to this case. The isolated horizon boundary conditions
imply that the intrinsic geometry of the quantum horizon is
described by the so called $\U(1)$ Chern-Simons theory on the
horizon. This is a well-developed, topological quantum field
theory. A deeply satisfying feature of the analysis is that there
is a seamless matching of three otherwise independent structures:
the isolated horizon boundary conditions which come from classical
general relativity; the quantum geometry in the bulk; and the
Chern-Simons theory on the horizon. In particular, one can
calculate eigenvalues of certain physically interesting operators
using purely bulk quantum geometry without any knowledge of the
Chern-Simons theory, or using the Chern-Simons theory without any
knowledge of the bulk quantum geometry. The two theories have
never heard of each other. Yet, thanks to the isolated horizon
boundary conditions, the two infinite sets of numbers match
exactly, providing a coherent description of the quantum horizon.

In this description, the polymer excitations of the bulk geometry,
each labelled by a spin $j_I$, pierce the horizon, endowing it an
elementary area $a_{j_I}$ given by (\ref{2.1}). The sum
$\textstyle{\sum_I} a_{j_I}$ adds up to the total horizon area
$a_{\rm hor}$. The intrinsic geometry of the horizon is flat
except at these puncture, but at each puncture there is a
\textit{quantized} deficit angle. These add up to endow the
horizon with a 2-sphere topology, as required by a quantum analog
of the Gauss-Bonnet theorem. For a solar mass black hole, a
typical horizon state would have $10^{77}$ punctures, each
contributing a tiny deficit angle. So, although the quantum
geometry \textit{is} distributional, it can be well approximated
by a smooth metric.

The counting of states can be carried out as follows. First one
constructs a micro-canonical ensemble by restricting oneself only
to those states for which the total area, angular momentum, and
charges lie in small intervals around fixed values $a_{\rm hor},
J_{\rm hor}, Q^i_{\rm hor}$. (As is usual in statistical
mechanics, the leading contribution to the entropy is independent
of the precise choice of these small intervals.) For each set of
punctures, one can compute the dimension of the surface Hilbert
space, consisting of Chern-Simons states compatible with that set.
One allows all possible sets of punctures (by varying both the
spin labels and the number of punctures), subject to the
constraint that the total area $a_{\rm hor}$ be fixed, and adds up
the dimensions of the corresponding surface Hilbert spaces to
obtain the number ${\cal N}$ of permissible surface states. One
finds that the horizon entropy $S_{\rm hor}$ is given by
\be \label{3.3} S_{\rm hor} := \ln {\cal N} =
\frac{\gamma_o}{\gamma}\, \frac{a_{\rm hor}}{\lp^2} + {\cal O}
(\frac{\lp^2}{a_{\rm hor}}) ,\quad {\rm where} \quad \gamma_o =
\frac{\ln 2}{\sqrt{3} \pi } \ee
Thus, for large black holes, entropy is indeed proportional to the
horizon area. This is a non-trivial result; for examples, early
calculations often led to proportionality to the square-root of
the area. However, even for large black holes, one obtains
agreement with the Hawking-Bekenstein formula \textit{only} in the
sector of quantum geometry in which the Barbero-Immirzi parameter
$\gamma$ takes the value $\gamma = \gamma_o$. Thus, while all
$\gamma$ sectors are equivalent classically, the standard quantum
field theory in curved space-times is recovered in the
semi-classical theory only in the $\gamma_o$ sector of quantum
geometry. It is quite remarkable that thermodynamic considerations
involving \textit{large} black holes can be used to fix the
quantization ambiguity which dictates such Planck scale properties
as eigenvalues of geometric operators. Note however that the value
of $\gamma$ can be fixed by demanding agreement with the
semi-classical result just in one case ---e.g., a spherical
horizon with zero charge, or a cosmological horizon in the de
Sitter space-time, or, \ldots. Once the value of $\gamma$ is
fixed, the theory is completely fixed and we can ask: Does this
theory yield the Hawking-Bekenstein value of entropy of
\textit{all} isolated horizons, irrespective of the values of
charges, angular momentum, and cosmological constant, the amount
of distortion, or hair. The answer is in the affirmative. Thus,
the agreement with quantum field theory in curved space-times
holds in \textit{all} these diverse cases.

Why does $\gamma_o$ not depend on other horizon parameters such as
the charges $Q^i_{\rm hor}$? This important property can be traced
back to a key consequence of the isolated horizon boundary
conditions: detailed calculations show that only the gravitational
part of the symplectic structure has a surface term at the
horizon; the matter symplectic structures have only volume terms.
(Furthermore, the gravitational surface term is insensitive to the
value of the cosmological constant.) Consequently, in the
geometric quantization procedure used in this analysis, there are
no independent surface quantum states associated with matter. This
provides a natural explanation of the fact that the
Hawking-Bekenstein entropy depends only on the horizon geometry
and is independent of electro-magnetic (or other) charges.

Finally, let us return to Wheeler's `It from Bit'. One can ask:
what are the states that dominate the counting? Perhaps not
surprisingly, they turn out to be the ones which assign to each
puncture the smallest quantum of area (i.e., spin value $j =
\textstyle{1\over 2}$), thereby maximizing the number of
punctures. In these states, each puncture defines one of Wheeler's
`elementary cell' and his two states correspond to the $j_z = \pm
1/2$ states, i.e. to whether the deficit angle is positive or
negative. However, in the complete theory, all values of $j$ (and
hence of $j_z$) must be allowed to obtain a complete description
of the geometry of the quantum horizon. If one is only interested
in counting states for large black holes, however, the leading
contribution comes from the $j = 1/2$ states.

To summarize, quantum geometry naturally provides the micro-states
responsible for the huge entropy associated with horizons. In this
analysis, all black holes ---including the ones of direct
astrophysical interest--- and cosmological horizons are treated in
an unified fashion. The sub-leading term has also been calculated
and shown to be proportional to $\ln a_{\rm hor}$ \cite{bh}.

\section{Conclusion}
\label{s4}

In this brief report, I have summarized two of the recent advances
which have answered some of the long standing questions of quantum
gravity raised in the Introduction. There have been two other
notable advances: i) the development of `spin-foam models'
discussed briefly by Baez at the conference which provide a new,
non-perturbative path integral approach to quantum gravity and
have led to a variety of interesting and intriguing mathematical
results on state sum models, extending in certain ways the very
interesting work on Turiev and Viro in 3-dimensions; and, ii) the
introduction of new measures on the space of connections relating
the quantum geometry framework to the standard Fock description of
photons and gravitons, which is paving the way to relate the
Planck scale calculations of quantum geometry to the more familiar
world of `low energy physics'. The vitality of the program is
reflected in the fact that many of the key ideas in all these
developments came from young researchers ---from Bojowald in
quantum cosmology, Krasnov in the understanding of quantum black
holes, Perez in spin foams and Varadarajan in the relation to low
energy physics.

Throughout the development of loop quantum gravity, unforeseen
simplifications have arisen regularly, leading to surprising
solutions to seemingly impossible difficulties. Progress could
occur because some of the obstinate problems which had slowed
developments in background independent approaches, sometimes for
decades, evaporated when `right' perspectives were found. I will
conclude with a few examples.

$\bullet$ Up until the early nineties, it was widely believed that
spaces of connections do not admit non-trivial diffeomorphism
invariant measures. This would have made it impossible to develop
our background independent approach. Quite surprisingly, such a
measure could be found by looking at connections in a slightly
more general perspective. It is simple, natural, and has just the
right structure to support quantum geometry. This geometry, in
turn, supplied some missing links, e.g., by providing just the
right expressions that Ponzano-Regge had to postulate without
justification in their celebrated, early work on 3-dimensional
gravity.

$\bullet$ Fundamental discreteness first appeared in a startling
fashion in the construction of the so-called `weaves', quantum
states which approximate given classical 3-geometries. In this
construction, quantum states based on finite graphs were
introduced as a starting point, with the goal of taking a
`continuum limit' as in lattice gauge theories. It came as a major
surprise that, if one wants to recover a given classical geometry
\textit{on large scales}, one can not take this limit, i.e., one
can not refine the underlying graph arbitrarily; there is an
in-built discreteness.

$\bullet$ Traces of holonomies of a suitably defined connection
around a smooth loop define a natural set of functions of
connections. At a heuristic level, it was found that they
automatically solve the most difficult part of the quantum
Einstein's equation. No one expected to find such simple and
natural solutions even heuristically. This calculation suggested
that the action of this part of the quantum Einstein equation is
concentrated at vertices of graphs, which in turn led to
strategies for its regularization.

$\bullet$ As I indicated in some detail, unforeseen insights arose
in the well-studied subject of quantum cosmology essentially by
taking an adequate account of the quantum nature of geometry,
i.e., by respecting the fundamental discreteness of the
eigenvalues of the scale factor operator. Similarly, in the case
of black holes, three quite distinct structures ---the isolated
horizon boundary conditions, the bulk quantum geometry and the
surface Chern-Simons theory--- blended together unexpectedly to
provide a coherent theory of quantum horizons.

Repeated occurrence of such `unreasonable' simplifications suggest
that the ideas underlying loop quantum gravity may have captured
an essential germ of truth.

\bigskip

\textbf{Acknowledgements} I would like to thank John Baez, Martin
Bojowald, Rodolfo Gambini, Jerzy Lewandowski, Alejandro Perez,
Jorge Pullin, Carlo Rovelli and Thomas Thiemann for numerous
discussions. This work was supported in part by the NSF grant
PHY-0090091 and the Eberly research funds of Penn State.

\end{document}